\documentclass[twocolumn,superscriptaddress,pre]{revtex4-1}
\usepackage{graphicx}
\usepackage{array}
\usepackage{amssymb}
\usepackage{amsfonts}
\usepackage{amsmath}
\usepackage{mathrsfs}
\usepackage{color}
\usepackage{booktabs}
\usepackage{threeparttable}
\usepackage{multirow}
\usepackage{subfigure}
\usepackage{times}
\usepackage{epsfig}
\usepackage{chngpage}
\usepackage{latexsym}
\usepackage{mathrsfs}
\usepackage{bm}

\begin{document}

\title{A universal data based method for reconstructing complex networks with
binary-state dynamics}

\author{Jingwen Li}
\affiliation{School of Systems Science, Beijing Normal University,
Beijing, 100875, China}

\author{Zhesi Shen}
\affiliation{School of Systems Science, Beijing Normal University,
Beijing, 100875, China}

\author{Wen-Xu Wang} \email{wenxuwang@bnu.edu.cn}
\affiliation{School of Systems Science, Beijing Normal University,
Beijing, 100875, China}
\affiliation{Business School, University of Shanghai for Science
and Technology, Shanghai 200093, China}

\author{Celso Grebogi}
\affiliation{Institute for Complex Systems and Mathematical Biology,
King's College, University of Aberdeen, Aberdeen AB24 3UE, UK}

\author{Ying-Cheng Lai}
\affiliation{Institute for Complex Systems and Mathematical Biology,
King's College, University of Aberdeen, Aberdeen AB24 3UE, UK}
\affiliation{School of Electrical, Computer and Energy Engineering,
Arizona State University, Tempe, Arizona 85287, USA}
\affiliation{Department of Physics, Arizona State University, Tempe,
Arizona 85287, USA}

\begin{abstract}

To understand, predict, and control complex networked systems, a
prerequisite is to reconstruct the network structure from observable data.
Despite recent progress in network reconstruction, binary-state dynamics that are ubiquitous in nature, technology and society still present an
outstanding challenge in this field. Here we offer a framework for reconstructing complex networks with binary-state dynamics by
developing a universal data-based linearization approach that is applicable to systems with linear, nonlinear, discontinuous, or stochastic dynamics governed by monotonous functions. The linearization procedure enables us to convert the network reconstruction into a sparse signal reconstruction problem that can be resolved through convex optimization.
We demonstrate generally high reconstruction accuracy for a number of complex networks
associated with distinct binary-state dynamics from using binary data contaminated by noise and missing data.
Our framework is completely data driven, efficient and robust, and does not require any a priori knowledge about the detailed dynamical process on the network. The framework represents a general paradigm for reconstructing, understanding, and
exploiting complex networked systems with binary-state dynamics.




\end{abstract}

\maketitle

\section{Introduction} \label{sec:intro}

Complex networked systems consisting of units with binary-state
dynamics are common in nature, technology, and society~\cite{barrat2008}.
In such a system, each unit can be in one of the two possible states,
e.g., being active or inactive in neuronal and gene regulatory
networks~\cite{kumar2010}, cooperation or defection in
networks hosting evolutionary game dynamics~\cite{game}, being
susceptible or infected in epidemic spreading on social and technological
networks~\cite{pastor2015}, two competing opinions in social
communities~\cite{shao2009}, etc. The interactions among the units
are complex and a state change can be triggered either
deterministically (e.g., depending on the states of their neighbors)
or randomly. Indeed, deterministic and stochastic state changes
can account for a variety of emergent phenomena, such as the outbreak
of epidemic spreading~\cite{granell2013}, cooperation among selfish
individuals~\cite{santos2005}, oscillations in biological
systems~\cite{koseska2013}, power blackout~\cite{buldyrev2010},
financial crisis~\cite{galbiati2013}, and phase transitions in natural
systems~\cite{balcan2011}. A variety of models have been introduced
to gain insights into binary-state dynamics on complex
networks~\cite{newman2010}, such as the voter models
for competition of two opinions~\cite{voter}, stochastic propagation
models for epidemic spreading~\cite{sis}, models of rumor diffusion
and adoption of new technologies~\cite{castellano2009}, cascading
failure models~\cite{bashan2013}, Ising spin models for ferromagnetic
phase transition~\cite{ising}, and evolutionary games for cooperation
and altruism~\cite{santos2008}. A general theoretical approach to dealing
with networks hosting binary state dynamics was developed
recently~\cite{gleeson2013} based on pair approximation and
master equations, providing a good understanding of the effect of
the network structure on the emergent phenomena.

In this paper, we address the inverse problem of binary-state dynamics
on complex networks, i.e., the problem of reconstructing the network
structure and binary dynamics from data. Deciphering the network structure
from data has always been a fundamental problem in complexity science, as
the structure can determine the type of collective dynamics on the
network~\cite{boccaletti2006}. More generally, for a complex networked
system, reductionism is not effective and it is necessary to reconstruct
and study the system as a whole~\cite{barabasi2011}.
The importance of network reconstruction has been increasingly recognized
and effective methodologies have been developed~\cite{GdiBLC:2003,friedman2004,
Timme:2007,CMN:2008,guo2008,RWLL:2010,HKKN:2011,WLGY:2011,barzel2013,
feizi2013,CCPGP:2013,SWFDL:2014,han2015}. Of particular relevance
to our work is spreading dynamics on complex networks, where the
available data are binary: a node is either infected or healthy.
In such cases, a recent work~\cite{SWFDL:2014} demonstrated that
the propagation network structure can be reconstructed
and the sources of spreading can be detected by exploiting compressive
sensing~\cite{CRT:2006a,CRT:2006b,Donoho:2006,Baraniuk:2007,CW:2008,
Romberg:2008}. However, for binary state network dynamics, a general
reconstruction framework was lacking (prior to the present work). The
problem of reconstructing complex networks with binary-state dynamics
is extremely challenging, for the following reasons. (i) The switching
probability of a node depends on the states of its neighbors according
to a variety of functions for different systems, which can be linear,
nonlinear, piecewise, or stochastic. If the function that governs the
switching probability is unknown, a tremendous difficulty would arise
in obtaining a solution of the reconstruction problem. (ii) Structural
information is often hidden in the binary states of the nodes in an
unknown manner and the dimension of the solution space can be extremely
high, rendering impractical (computationally prohibitive) brute-force
enumeration of all possible network configurations. (iii) The presence
of measurement noise, missing data, and stochastic effects in the
switching probability make the reconstruction task even more challenging,
calling for the development of effective methods that are robust
against internal and external random effects.

To meet the challenges, we develop a general and robust framework
for reconstructing complex networks based solely on the binary
states of the nodes without any knowledge about the switching
functions. Our idea is centered around developing a general method
to linearize the switching functions from binary data. The data-based
linearization method is applicable to linear, nonlinear, piecewise, or
stochastic switching functions. The method allows us to convert the network
reconstruction problem into a sparse signal reconstruction problem for local
structures associated with each node. Exploiting the natural sparsity of complex networks, we
employ the lasso~\cite{lasso}, an L$_1$ constrained fitting method
for statistics and data mining, to identify the neighbors of each node
in the network from sparse binary data contaminated by noise. We
establish the underlying mechanism that justifies the linearization
procedure by conducting tests using a number of linear, nonlinear and
piecewise binary-state dynamics on a large number of model and real
complex networks. We find universally high reconstruction accuracy
even for small data amount with noise. Because of its high accuracy,
efficiency and robustness against noise and missing data,
our framework is promising as a general solution to the inverse
problem of network reconstruction from binary-state time series, which is
key to articulating effective strategies to control complex networks with
binary state dynamics using, e.g., the recently developed network
controllability frameworks~\cite{liu2011,nepusz2012,yan2012,yuan2013,
RR:2014,Wuchty:2014}. The data-based linearization method is also useful
for dealing with general nonlinear systems with a wide range of applications.

\section{Binary-state dynamics} \label{sec:binary}

We consider a large number of representative binary state processes on
complex networks, which model a plethora of physical, social and biological
phenomena~\cite{gleeson2013}. In such a dynamical process, the state
of a node can be $0$ (inactive) or $1$ (active). In general, the
process can be characterized by two switching functions, $F(m,k)$ and
$R(m,k)$, which determine the probabilities for a node to change its
state from $0$ to $1$ and vice versa, respectively. The variables in
these functions, $k$ and $m$, are the degree of the node and the
number of active neighbors of the node, respectively. The switching
functions can be linear, nonlinear, piecewise, bounded and stochastic
for characterizing and generating all kinds of binary-sate dynamical
processes occurring on complex networks.
Despite the difference among the switching functions, the feature
that a node's switching probability depends on its degree and its
number of active neighbors is generic. Table~\ref{tab:dynamics}
lists the switching functions of different models, and the brief descriptions
of each model can be found in Appendix.

\begin{table}[h]
\linespread{1.3}
\renewcommand{\arraystretch}{1.3}
\addtolength{\tabcolsep}{0.1pt}
\centering{}
\caption{{\bf Switching functions for various binary state dynamical
processes on complex networks.} The function $F(m,k)$ is the
probability that a node switches its state from $0$ to $1$ while
$R(m,k)$ represents the probability of the reverse process, where
$k$ is the degree of the node, $m$ is the number of neighbors of
this node in the active state $1$. The models and the other
parameters are described in {\bf Methods}. The parameter values
used in the simulations are listed in Supplementary Table~S1.}
\scriptsize{
\begin{tabular}{c c c}
\hline
Model & $F(m,k)$ & $R(m,k)$\\
\hline
Voter~\cite{voter} & $\frac{m}{k}$ & $\frac{k-m}{k}$\\
Kirman~\cite{kirman} & $c_1+dm$ & $c_2+d(k-m)$\\
Ising Glauber~\cite{ising,glauber} & $\dfrac{1}{1+e^{\frac{\beta}{k}(k-2m)}}$ & $\dfrac{e^{\frac{\beta}{k}(k-2m)}}{1+e^{\frac{\beta}{k}(k-2m)}}$\\
SIS~\cite{sis} & $1-(1-\lambda)^m$ & $\mu$\\
Game~\cite{game} & $\dfrac{1}{\alpha+e^{\frac{\beta}{k} [(a-c)(k-m)+(b-d)m]}}$ & $\dfrac{1}{\alpha+e^{\frac{\beta}{k} [(c-a)(k-m)+(d-b)m]}}$\\
Language~\cite{language} & $s(\frac{m}{k})^\alpha$ & $(1-s)(\frac{k-m}{k})^\alpha$\\
Threshold~\cite{threshold}
& $\begin{cases}
0& \text{if } m\leqslant M_k\\
1& \text{if } m>M_k
\end{cases}$
& $0$ \\
Majority vote~\cite{majority}
& $\begin{cases}
Q& \text{if } m<k/2\\
1/2& \text{if } m=k/2\\
1-Q& \text{if } m>k/2
\end{cases}$
& $\begin{cases}
1-Q& \text{if } m<k/2\\
1/2& \text{if } m=k/2\\
Q& \text{if } m>k/2
\end{cases}$ \\
\hline
\end{tabular}
\label{tab:dynamics}
}
\end{table}

\section{Reconstruction method} \label{sec:methods}

Our goal is to articulate a general framework to reconstruct the network structure from binary
states of nodes without knowing {\em a priori} the specific
switching functions. A key step is to develop a universal
procedure to obtain the linearization of the switching functions
from binary data. We demonstrate that this can be accomplished by
taking advantage of certain common features of the binary
state dynamics.

\begin{figure*}
\centering
\includegraphics[width=0.9\textwidth]{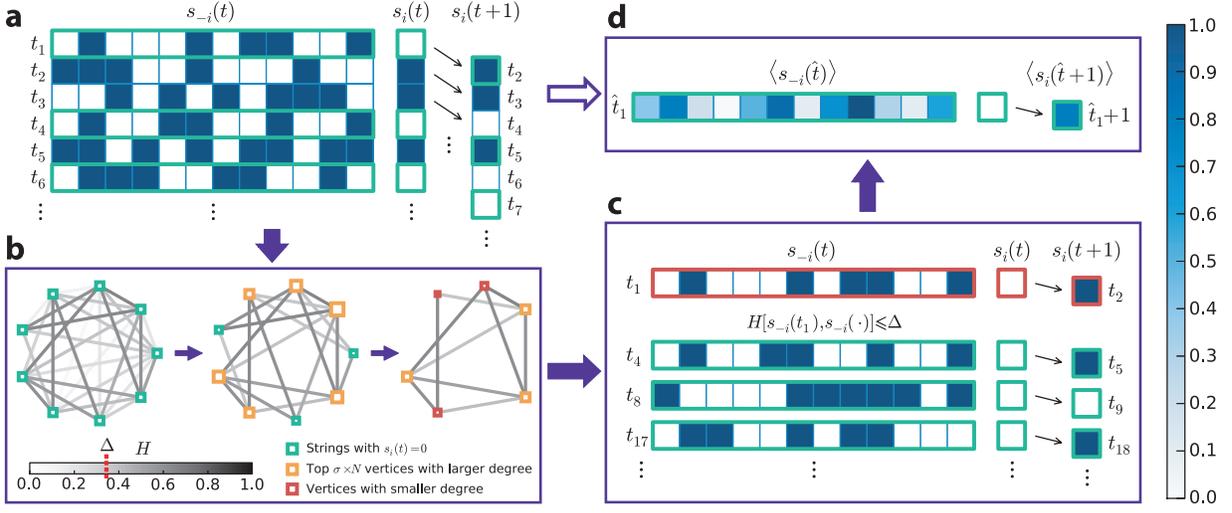}
\caption{{\bf Schematic illustration of data based linearization
from a merging process.} ({\bf a}) The original binary-state time series,
where the dark blue squares denote the $1$ state and the white squares denote
the $0$ state. The variable $s_{-i}(t)$ consists of $s_j(t)$ for all
$j \neq i$. Only strings with $s_i(t)=0$ (highlighted by the green frames)
contain useful information for reconstruction. We identify the time steps
with $s_i(t)=0$ and use $s_i(t+1)$. ({\bf b}) Method of choosing bases.
We first construct a network where the vertices denote strings of
$s_{-i}(t)$ with $s_i(t)=0$ (green squares) and the edges are weighted
by the normalized Hamming distance $H$ between the strings. We then eliminate
the edges with weight smaller than the threshold $\Delta$. Setting another
threshold $\sigma$, we obtain the top $\sigma \times N$ vertices with
large degrees (yellow squares) and remove the other vertices together
with their edges. Finally, we pick out the vertices with smaller degree
(red squares) according to the number of base strings needed
for reconstruction. ({\bf c}) Selection of subordinate strings from
a base.
We take $t_1$ as a base $\hat t_1$ and calculate $H$ between
$s_{-i}(t_1)$ and other strings $s_{-i}(t)$ so as to sort out the time
steps satisfying $H[s_{-i}(t_1),s_{-i}(\cdot)]<\Delta$ in this set.
({\bf d}) Establishing average node states. We calculate the average
value $\left\langle s_{-i}(\hat t )\right\rangle$ to represent the
state of the data set subject to the base, and the average value
$\left\langle s_i(\hat t +1)\right\rangle$ to linearize the switching
probability $P_i^{01}(t)$ [Eqs.~(\ref{eq:linearize})]. The average values are shown in blue. Similarly,
we obtain a sequence of $\hat{t}_M$ and the associated average values
for reconstructing network structure by employing the lasso to solve
${\bf Y}_i=\Phi_i\times {\bf X}_i$ (see {\bf Methods} for details).}
\label{fig1}
\end{figure*}

\subsection{Data based linearization of switching functions}

To proceed, we note that the number of active neighbors at time
$t$ can be expressed as
\begin{equation}
m_i(t)=\sum^N_{j=1,j\neq i}a_{ij}s_j(t) \text{,}
\end{equation}
where $a_{ij}=1$ if nodes $i$ and $j$ are connected and $a_{ij}=0$
otherwise, and $s_j(t)$ denotes the state of node $j$ at time step
$t$. In general, the switching probability $P_i^{01}(t)$ for node $i$ to
change its state from $0$ to $1$ at time step $t$ can be written as
\begin{equation} \label{eq:nonlinear}
P_i^{01}(t) = F\left( m_i(t),k_i\right)=
F\left( \sum^N_{j=1,j\neq i}a_{ij}s_j(t), k_i\right),
\end{equation}
where $F$ is a monotonous function characterizing different
dynamical models, e.g., those listed in Table~\ref{tab:dynamics}.
In Eq.~(\ref{eq:nonlinear}), all the matrix elements $a_{ij}$
($i,j=1,\ldots,N$) that are to be inferred from data characterize the network structure. In general this is a difficult problem, because
in Eq.~(\ref{eq:nonlinear}), only nodal state $s_j(t)$ is measurable,
whereas neither of the quantities $k_i$ and $P_i^{01}(t)$ nor the
form of $F$ is known. In fact, not knowing the function $F$ is
the main difficulty in reconstructing the adjacency matrix
$\{a_{ij}\}$. To overcome this difficulty, we propose a
merging process to linearize $F$, i.e.,
\begin{equation}
F \sim c_i\cdot\sum^N_{j=1,j\neq i}a_{ij}s_j(t)+d_i,
\end{equation}
where $c_i$ and $d_i$ are constants associated with node $i$. Insofar
as the linearization is realized, we can solve $a_{ij}$. The idea of linearization is first proposed and used in Ref.~\cite{SWFDL:2014}, but the mathematical form of $F$ is assumed to be known in that case. It is worth
noting that the linearization approach is highly nontrivial and is
fundamentally different from that in the standard canonical nonlinear
analysis because, in our case, the mathematical form of $F$ is
not available, which can be a nonlinear, discrete and piecewise function.
The fully data based linearization procedure is the main contribution
of this paper.

\subsection{Procedure of dealing with binary-state data}
We present the procedure of dealing with binary-state data.
The merging based linearization process enables the
probability $P_i^{01}(t)$ to be estimated according to the law of
large numbers, from which the solution of $a_{ij}$ can be obtained.
In particular, as shown in Fig.~\ref{fig1}(a), for an arbitrary node $i$, we first identify all
the time steps with $s_i(t)=0$ as information about the switching
probability $P_i^{01}(t)$ is contained only in the flipping behavior
from state 0.
Then we propose a method to choose the optimal base strings
that are neither too special nor too similar to each other(see Fig.~\ref{fig1}(b)).
A base string at $\hat{t}$ is a state vector $s_{-i}(\hat{t})$ based on which a set of similar
strings are identified and averaged to estimate the swiching probability.
Specifically, we first construct a network whose vertices
represent strings composed of $s_j(t) (j\neq i)$ at different
time steps for $s_i(t)=0$ and the edges are weighted by the
normalized pairwise Hamming distances among the strings. We
then eliminate edges whose weight is smaller than a threshold,
say $\Delta$.
Setting another threshold $\sigma$,
we can extract a subnetwork where only the top $\sigma \times N$
vertices of large degree are preserved, while other vertices
and their edges are removed.
Finally, we pick out $M$ vertices with smallest degrees ensuring that the selected base strings are
sufficiently different, where $M$ is the number of equations in Eq.~(\ref{eq:matrix})
For each chosen base string, we set a
threshold $\Delta$ in the normalized Hamming distance between strings
to select a set of subordinate strings that belong to each base string, as shown in Fig.~\ref{fig1}(c).
A subordinate string is a string whose normalized Hamming distance to the base string is less than the selected threshold $\Delta$.
Using the average of $s_j(t)$ to represent the state of node $j$ and the average of
$s_i(t+1)$ to estimate the switching probability $P_i^{01}(t)$ of
node $i$ according to the law of large numbers, we obtain
$P_i^{01}(t) \approx \langle s_i(\hat t +1)\rangle$, where $\hat t$
denotes the time of the base string (see Fig.~\ref{fig1}(d)).

The whole process leads to the linearization of $F$ with the following data-based relationship
\begin{equation}
\label{eq:linearize}
\langle s_i(\hat t +1)\rangle \approx c_i\cdot
\sum^N_{j=1,j\neq i}a_{ij}\langle s_j(\hat t)\rangle+d_i \text{,}
\end{equation}
where $\langle\cdot\rangle$ is the average over all time $t$ of the
subordinate strings within $\hat t$. The constant parameter $k_i$ is
incorporated into the linear coefficient $c_i$ and the intercept $d_i$.
It is not necessary to estimate the quantities $c_i$, $a_{ij}$ and $d_i$
in Eq.~(\ref{eq:linearize}) separately - it is only necessary to infer
value of the product $c_i\times a_{ij}$. In particular, if $i$ and $j$
are not connected, we have $c_i\times a_{ij}=0$, but a nonzero value
of $c_i\times a_{ij}$ means that there is a link between the two nodes.
As we will show, the value of $d_i$ can be obtained but this quantity
plays little role in the reconstruction.

\begin{figure*}
\centering
\includegraphics[width=0.8\textwidth]{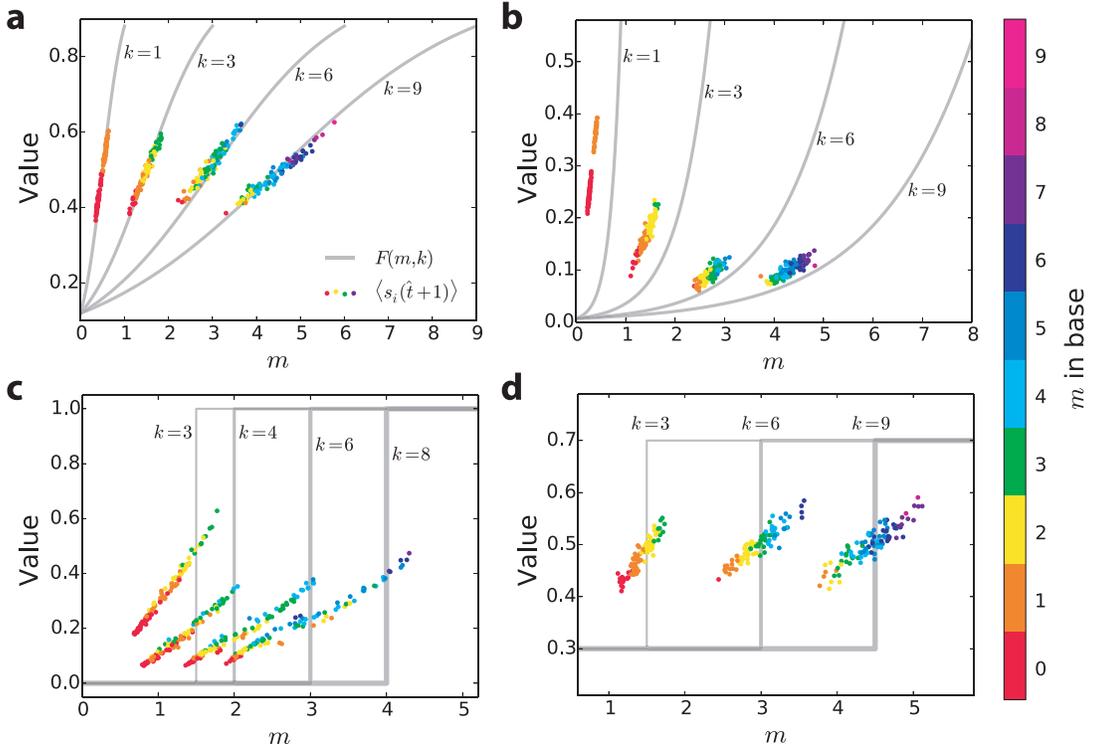}
\caption{{\bf Data based linearization procedure for nonlinear and
piecewise binary-state dynamics.} Linearization of the switching
probability function $F(m,k)$ for ({\bf a}) Ising model, ({\bf b}) evolutionary
game model, ({\bf c}) threshold model, and ({\bf d}) majority model. The grey lines
represent Eq.~(\ref{eq:nonlinear}) with the function $F(m,k)$ from the
different models, where $k$ is the node's degree and $m$ is the number of
active neighbors. Data points are the result of the linearization
procedure from time series, which corresponds to Eq.~(\ref{eq:linearize}).
For the linearized function, $m$ is obtained from
$\sum^N_{j=1,j\neq i}a_{ij}\langle s_j(\hat t)\rangle$ and the value
of the function is obtained via $\left\langle s_i(\hat t +1)\right\rangle$.
For the data points, each color represents a set of subordinate strings
whose base string has $m$ active neighbors. The colors
demonstrate that bases with different $m$ values are needed
to produce a linear function with a sufficient range of $m$ for
reconstruction, which justifies the base selection based on the
normalized Hamming distance in Fig.~\ref{fig1}. For both nonlinear
and piecewise switching functions, a linearized function in the form
of Eq.~\eqref{eq:linearize} can be generated based entirely on data,
which is the key to reconstruction. The data points
are obtained from an ER random network of $N=100$ nodes and average
degree $\langle k\rangle=6$.}
\label{fig2}
\end{figure*}

Figure~\ref{fig2} shows some representative examples to validate the
linearization procedure. Four types of dynamics, including two with
continuous and nonlinear switching functions and two with discontinuous
and piecewise functions, are tested. We see that the switching functions
$F$ for different parameter values are linearized, enabling the network
structure in the linearized system~(\ref{eq:linearize}) to be reconstructed
by distinguishing between zero and nonzero values of the reconstructed
product $c_i\times a_{ij}$. As compared to the original function $F$,
the range of $m$ in the linearized function typically shrinks considerably as
a result of the merging process, as shown in Figs.~\ref{fig2}(a) and \ref{fig2}(b).
For the discrete piecewise functions in Figs.~\ref{fig2}(c) and
\ref{fig2}(d), approximately linear functions arise for different parameter values. This is particularly striking, because even given a switching function, it is still difficult to linearize a piecewise function. We have achieve a data-based linearization of nonlinear and piecewise functions without any knowledge a priori.

\subsection{Theoretical validation of data-based linearization}
We provide an analysis for the completely data-based linearization that gives rise to the general relationship (Eq.~(\ref{eq:linearize})) from general binary-state dynamics characterized by the switching probability (Eq.~(\ref{eq:nonlinear})),

For nodes with only one neighbor, the linear relationship~(\ref{eq:linearize}) can be rigorously proved. In this scenario, the number of active neighbors is either $0$ or $1$. Let $P_{\hat t}(1)$ denote the proportion of strings with single active neighbors in the set of base $\hat t$, and denote the proportion of strings with null active neighbors as $1-P_{\hat t}(1)$. Let the switching probability of null active neighbors and single active neighbors be $f(0)$ and $f(1)$. Then we have
\begin{eqnarray}
\langle s_i(\hat t +1)\rangle \approx \langle P_i^{01}(t)\rangle &=& f(0)\left[ 1-P_{\hat t}(1)\right] +f(1)P_{\hat t}(1) \nonumber \\
 &=& \left[f(1)-f(0)\right] P_{\hat t}(1)+f(0)
\label{eq:k1_y}
\end{eqnarray}
and
\begin{equation}
\sum^N_{j=1,j\neq i}a_{ij}\langle s_j(\hat t )\rangle = P_{\hat t}(1) \text{.}
\label{eq:k1_x}
\end{equation}
Inserting Eq.~(\ref{eq:k1_x}) into Eq.~(\ref{eq:k1_y}), we have
\begin{equation}
\langle s_i(\hat t +1)\rangle \approx \left[f(1)-f(0)\right]\sum^N_{j=1,j\neq i}a_{ij}\langle s_j(\hat t )\rangle +f(0) \text{,}
\label{eq:linear_k1}
\end{equation}
which is a linear form that is subject to Eq.~(\ref{eq:linearize}), because both $[f(1)-f(0)]$ and $f(0)$ are constants and they are determined by the specific binary-state dynamics.

\begin{figure*}
\centering
\includegraphics[width=0.9\textwidth]{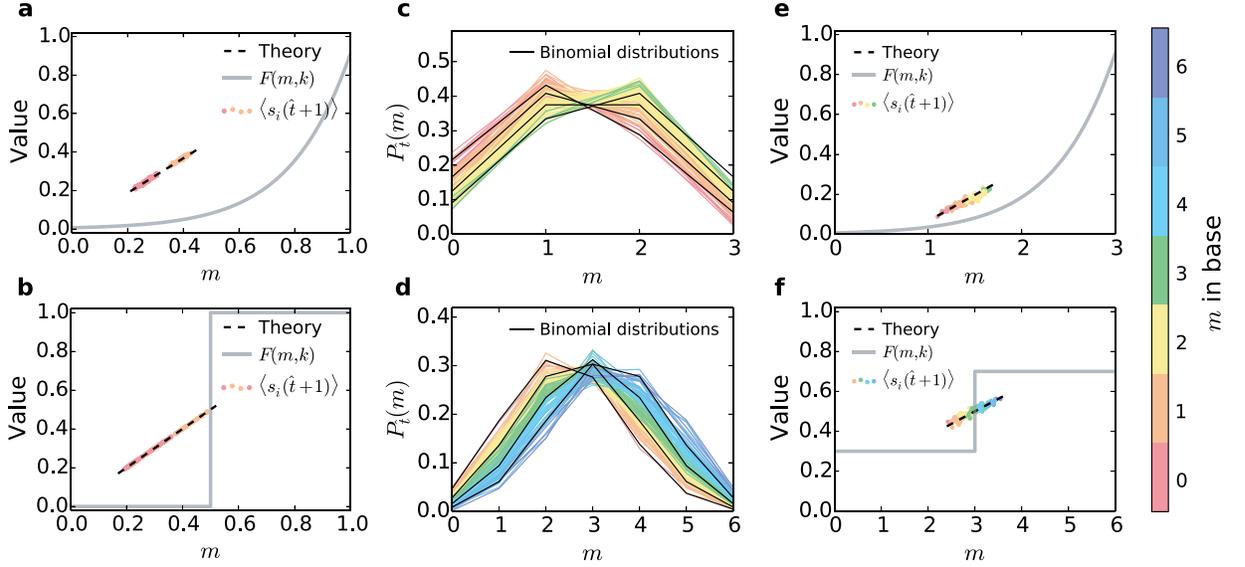}
\caption{{\bf Theoretical analysis of the data-based linearization.}
({\bf a, b})Linearization of switching function for nodes with a single neighbor for the game model ({\bf a}) and the threshold model ({\bf b}). The grey solid curves are the original switching functions, data points are the results of data-based linearization (Eq,~(\ref{eq:linearize})), and the dashed lines are theoretical predictions from Eq.~(\ref{eq:linear_k1}).
The color of data points represents two sets of subordinate strings whose base string has no active neighbors ($m=0$) or has a single active neighbors ($m=1$).For both nonlinear and piecewise switching functions, the theoretical predictions are in exact agreement with data-based linearization, because for $k_i=1$ the linearization is rigorous without any approximation.
({\bf c, d}) The distribution of active neighbors $m$ in subordinate strings subject to each base string and binomial distributions for reconstructing node $i$ with $k_i=3$ for the game model ({\bf c}) and $k_i=6$ for the majority model ({\bf d}), respectively.
Each color of curves represents a set of subordinate strings whose base string has $m$ active neighbors. The distribution can be well described by binomial distributions under different success probability in each trial, as exemplified by black curves. There is a good agreement between the distribution of active neighbors in subordinate strings and binomial distributions.
({\bf e, f}) The original switching function and the linearized function with theoretical prediction based on binomial distribution for the game model ({\bf e}) and the majority model ({\bf f}), respectively.
The color of data points represents different sets of subordinate strings whose base string has different number of active neighbors $m$ (the same meaning as in ({\bf c}), ({\bf d})).
The grey curves are the original switching function in the binary-state dynamics. The black dashed lines are the theoretical prediction of the linear relationship through Eq.~(\ref{eq:Taylor_linear}) based on binomial distribution and Taylor linear approximation. The theoretical predictions are in good agreement with numerical results. }
\label{fig:validation}
\end{figure*}

Figures~\ref{fig:validation}(a, b) shows two representative examples of reconstructing the local structure of a node with one neighbor for the evolutionary game model and the threshold model. We see explicitly linear relationship for both models. With respect to different number of active neighbors in the original bases, two sets of groups are classified.


For nodes with more than one neighbor, the linear relationship can be justified and predicted based on binomial distribution and Taylor linear approximation. For an arbitrary node, say, node $i$ with $k$ neighbors, we will substantiate the linear relationship between $\langle s_i(\hat t +1)\rangle$ and $\sum^N_{j=1,j\neq i}a_{ij}\langle s_j(\hat t )\rangle$ resulting from the data-based linearization, where
\begin{eqnarray}
\langle s_i(\hat t +1)\rangle \approx \langle P_i^{01}(t)\rangle = \sum^{k_i}_{m=0}F(m,k_i)P_{\hat t}(m),
\label{eq:k_y}
\end{eqnarray}
and
\begin{equation}
\sum^N_{j=1,j\neq i}a_{ij}\langle s_j(\hat t )\rangle = \sum^{k_i}_{m=0}mP_{\hat t}(m) \text,
\label{eq:k_x}
\end{equation}
where $P_{\hat t}(m)$ represents the proportion of strings with $m$ active neighbors among all strings that belong to the set of base $\hat t$. The key to validating the linear relationship lies in the distribution that $P_{\hat t}(m)$ obeys.

Regarding the effect of the merging process as shown in Fig.~\ref{fig1}, we hypothesize that $P_{\hat t}(m)$ follows binomial distributions with different success probability $p_{\hat t}$. We denote the proportion of state $0$ in data to be $p_0$. If the strings are randomly chosen for each set of a base, $P_{\hat t}(m)$ exactly obeys binomial distribution with success probability $p_0$. However, due to the process of selecting strings that are similar to each set of a base, the distribution will be biased toward the number of active neighbors in the base. Despite the original complex influence of the base and string selections based on Hamming distance, their effects can be simply regarded as selecting a group of strings with similar proportion of state $0$ since we actually do not know which the node's neighbors are. This process leads to the success probability that depends on the base string.
Figs.~\ref{fig:validation}(c, d) shows the comparison between the actual distribution of $P_{\hat t}(m)$ obtained from numerical simulations and the binomial distributions with different success probability in each trial in the game and majority model, where the success probability in each trial approximately range from $0.4$ to $0.6$ because $p_0 \approx 0.5$ in the data. We see that $P_{\hat t}(m)$ can be well approximated by binomial distributions with different parameter values, which indeed validates our binomial distribution hypothesis.

Based on the binomial distribution hypothesis, we have
\begin{equation}
P_{\hat t}(m) = \mathrm{C}_{k_i}^m {p_{\hat t}}^m (1-p_{\hat t})^{k_i-m}.
\label{eq:binomial}
\end{equation}
Inserting Eq.~(\ref{eq:binomial}) into Eq.~(\ref{eq:k_y}) yields
\begin{small}
\begin{eqnarray}
\langle s_i(\hat t +1)\rangle
&\approx& \sum^{k_i}_{m=0}F(m,k_i)\mathrm{C}_{k_i}^m {p_{\hat t}}^m (1-p_{\hat t})^{k_i-m} \nonumber \\
&=& \sum^{k_i}_{m=0}\mathrm{C}_{k_i}^m \sum_{l=0}^{m}\left[ (-1)^{m-l}\mathrm{C}_{m}^lF(l,k_i)\right] {p_{\hat t}}^m.
\label{eq:Taylor_0}
\end{eqnarray}
\end{small}
The fact that $p_{\hat t}$ fluctuates around $p_0$ allows us to apply the Taylor series expansion around $p_0$ to Eq.~(\ref{eq:Taylor_0}), leading to
\begin{small}
\begin{eqnarray}
\langle s_i(\hat t +1)\rangle
&\approx& \sum^{k_i}_{m=0}\mathrm{C}_{k_i}^m \sum_{l=0}^{m}\left[ (-1)^{m-l}\mathrm{C}_{m}^lF(l,k_i)\right] p_0^m \nonumber \\
&+& \sum^{k_i}_{m=0}\mathrm{C}_{k_i}^m \sum_{l=0}^{m}\left[ (-1)^{m-l}\mathrm{C}_{m}^lF(l,k_i)\right] mp_0^{m-1}(p_{\hat t}-p_0) \nonumber \\
&+& \mathcal{O}(p_{\hat t}-p_0).
\end{eqnarray}
\end{small}
Omitting the high-order term $\mathcal{O}(p_{\hat t}-p_0)$, we have
\begin{small}
\begin{eqnarray}
\langle s_i(\hat t +1)\rangle
&\approx& \sum^{k_i}_{m=0}\mathrm{C}_{k_i}^m \sum_{l=0}^{m}\left[ (-1)^{m-l}\mathrm{C}_{m}^lF(l,k_i)\right] (1-m)p_0^m \nonumber \\
&+& \sum^{k_i}_{m=0}\mathrm{C}_{k_i}^m \sum_{l=0}^{m}\left[ (-1)^{m-l}\mathrm{C}_{m}^lF(l,k_i)\right] mp_0^{m-1}p_{\hat t}. \nonumber \\
\label{eq:Taylor_y}
\end{eqnarray}
\end{small}
On the other hand, substitute Eq.~(\ref{eq:binomial}) into Eq.~(\ref{eq:k_x}) yields
\begin{eqnarray}
\sum^N_{j=1,j\neq i}a_{ij}\langle s_j(\hat t )\rangle &=& \sum^{k_i}_{m=0}m\mathrm{C}_{k_i}^m {p_{\hat t}}^m (1-p_{\hat t})^{k_i-m} \nonumber \\
&=& k_i p_{\hat t} \text{.}
\label{eq:Taylor_x}
\end{eqnarray}
Combining Eq.~(\ref{eq:Taylor_y}) and Eq.~(\ref{eq:Taylor_x}), we have
\begin{small}
\begin{eqnarray}
\langle s_i(\hat t +1)\rangle
&\approx& \sum^{k_i}_{m=0}\mathrm{C}_{k_i}^m \sum_{l=0}^{m}\left[ (-1)^{m-l}\mathrm{C}_{m}^lF(l,k_i)\right] (1-m)p_0^m \nonumber \\
&+& \left\lbrace \frac{1}{k_i}\sum^{k_i}_{m=0}\mathrm{C}_{k_i}^m \sum_{l=0}^{m}\left[ (-1)^{m-l}\mathrm{C}_{m}^lF(l,k_i)\right] mp_0^{m-1}\right\rbrace \nonumber\\
&\times& \sum^N_{j=1,j\neq i}a_{ij}\langle s_j(\hat t )\rangle.
\label{eq:Taylor_linear}
\end{eqnarray}
\end{small}
Note that all variables in the first term on the right hand side of Eq.~(\ref{eq:Taylor_linear}) are only determined by the binary-state dynamics and the node degree of $i$. Hence, the first term corresponding to $d_i$ is a constant with respect to node state $s_i$. In analogy, all variables in the coefficient of the second term are determined by the binary-state dynamics and the node degree of $i$ as well, indicating the coefficient is a constant corresponding to $c_i$ in Eq.~(\ref{eq:linearize}). Taken together, we theoretically justified that Eq.~(\ref{eq:Taylor_linear}) is approximately a linear equation in the form of Eq.~(\ref{eq:linearize}).


Figures~\ref{fig:validation}(e, f) shows the relationship between $\langle s_i(\hat t +1)\rangle$ and $\sum^N_{j=1,j\neq i}a_{ij}\langle s_j(\hat t )\rangle$ (namely $\langle m\rangle$) of each set of bases and the linear relationship calculated by using Eq.~(\ref{eq:Taylor_linear}) for the game model and the majority model with nonlinear and piecewise switching dynamics.
We see that the theoretical predictions are in good agreement with the results from the merging process for linearization, which strongly validates the data-based linearization for general binary-state dynamics.

It is noteworthy that the key to the success of the data-based linearization lies in selecting similar strings subject to a base and the average over each set of bases.
The selection of similar strings accounts for the binomial distribution of active neighbors in a set, and different bases induces different success probability in each trial. Then the average of the binomial distributions leads to the relatively small range of $\langle m \rangle$ compared to the original range in the switching function, allowing us to use Taylor linear approximation. Moreover, high-order terms in the Taylor series expansion contribute little to the binomial distribution, which justifies the low-order approximation. Based on the linear relationship, the reconstruction of local structure can be realized by employing the lasso without requiring the linear coefficients and intercept. In other words, the data-based linearization is general valid for arbitrary binary-state dynamics without any knowledge of the switching function.


\subsection{Reconstruction of local structure based on the lasso}

The linear relationaship, Eq.~(\ref{eq:linearize}) allows us to ascertain the neighbors of
any node $i$ from $M$ different values of the base time, e.g.,
$\hat t_1, \cdots, \hat t_M$, and their subordinate times. In particular,
with respect to $\hat t_1, \cdots, \hat t_M$, Eq.~(\ref{eq:linearize})
can be expressed in the matrix form ${\bf Y}_i=\Phi_i\times {\bf X}_i$ as Eq.~(\ref{eq:matrix})
, where the vector ${\bf X}_i$ is to be solved for obtaining the neighbors of $i$, and the vector ${\bf Y}_i$
and the matrix $\Phi_i$ can be constructed entirely from binary time
series without requiring any other information.
\begin{widetext}
\begin{equation}
\label{eq:matrix}
\begin{bmatrix}
\left\langle s_i(\hat t_1+1)\right\rangle \\
\left\langle s_i(\hat t_2+1)\right\rangle \\
\vdots \\
\left\langle s_i(\hat t_M+1)\right\rangle \\
\end{bmatrix}
=
\begin{bmatrix}
1 & \left\langle s_1(\hat t_1)\right\rangle & \cdots & \left\langle s_{i-1}(\hat t_1)\right\rangle & \left\langle s_{i+1}(\hat t_1)\right\rangle & \cdots & \left\langle s_N(\hat t_1)\right\rangle \\
1 & \left\langle s_1(\hat t_2)\right\rangle & \cdots & \left\langle s_{i-1}(\hat t_2)\right\rangle & \left\langle s_{i+1}(\hat t_2)\right\rangle & \cdots & \left\langle s_N(\hat t_2)\right\rangle \\
\vdots & \vdots &\vdots &\vdots &\vdots &\vdots &\vdots \\
1 & \left\langle s_1(\hat t_M)\right\rangle & \cdots & \left\langle s_{i-1}(\hat t_M)\right\rangle & \left\langle s_{i+1}(\hat t_M)\right\rangle & \cdots & \left\langle s_N(\hat t_M)\right\rangle \\
\end{bmatrix}
\begin{bmatrix}
d_i \\
c_i\cdot a_{i1} \\
\vdots \\
c_i\cdot a_{i,i-1} \\
c_i\cdot a_{i,i+1} \\
\vdots \\
c_i\cdot a_{iN} \\
\end{bmatrix}.
\end{equation}
\end{widetext}


The natural sparsity of
complex networks ensures that, on average, the number of neighbors for
a node is much smaller than the network size $N$, implying that ${\bf X}_i$
is typically sparse with most of its elements being zero and the
number of nonzero elements is in fact the node degree $k_i$ with
$k_i \ll N$. We can then exploit the sparsity to reconstruct ${\bf X}_i$
by employing the lasso~\cite{lasso}, a convex optimization method for
sparse signal reconstruction. The lasso incorporating an L1-norm and
an error control term is efficient and robust, enabling a reliable
reconstruction of the local network structure as represented by
${\bf X}_i$ from a small amount of data.
In particular, the problem is to optimize
\begin{equation}
\min_{{\bf X}_i} \Bigl\lbrace \dfrac{1}{2M}\Vert\Phi_i{\bf X}_i- {\bf Y}_i\Vert_2^2 + \lambda\Vert{\bf X}_i\Vert_1 \Bigr\rbrace \text{,}
\label{eq:Lasso}
\end{equation}
where $\Vert{\bf X}_i\Vert_1=\sum^N_{j=1,j\neq i}\vert x_{ij}\vert$
is the $L_1$ norm of ${\bf X}_i$ assuring the sparsity of the solution,
and the least squares term $\Vert\Phi_i{\bf X}_i- {\bf Y}_i\Vert_2^2$
guarantees the robustness of the solution against noise in data.
In Eq.~\eqref{eq:Lasso}, $\lambda$ is a nonnegative regularization
parameter that affects the reconstruction performance in terms
of the sparsity of the network, which can be determined by a
cross-validation method~\cite{lasso_python}. An advantage of using
the lasso is that $M$, i.e., the number of bases needed, can be much
less than the length of ${\bf X}_i$. For each base of each node, the
strings included can be collected and calculated from only one set of
data sample in the time series, ensuring the sparse data requirement.

After the vector ${\bf X}_i$ has been reconstructed, the direct neighbors
of node $i$ are simply those associated with nonzero elements in
${\bf X}_i$. In the same manner, we can uncover the neighborhoods of
all other nodes, so that the full structure of the network can be
obtained by matching the neighbors of all nodes.





\section{Reconstruction performance} \label{sec:results}

\subsection{Measurement indices}
To quantify the performance of our reconstruction method, we introduce two standard measurement indices,
the area under the receiver operating characteristic curve (AUROC) and the area under the precision-recall curve (AUPR).
True positive rate (TPR), false positive rate (FPR), Precision and Recall that are used to calculate AUROC and AUPR are defined as follows:
\begin{equation}
\mathrm{TPR}(l)=\frac{\mathrm{TP}(l)}{P},
\end{equation}
where $l$ is the cutoff in the edge list, $\mathrm{TP}(l)$ is the number of true positives in the top $l$
predictions in the edge list, and $P$ is the number of positives in the gold standard.
\begin{equation}
\mathrm{FPR}(l)=\frac{\mathrm{FP}(l)}{Q},
\end{equation}
where $\mathrm{FP}(l)$ is the number of false positive in the top $l$ predictions in the edge list,
and $\mathrm{Q}$ is the number of negatives in the gold standard.

\begin{equation}
\mathrm{Precision}(l)=\frac{\mathrm{TP}(l)}{\mathrm{TP}(l)+\mathrm{FP}(l)}=\frac{\mathrm{TP}(l)}{l},
\end{equation}
\begin{equation}
\mathrm{Recall}(l)=\frac{\mathrm{TP}(l)}{P},
\end{equation}
where $\mathrm{Recall}(l)$, which is called sensitivity, is equivalent to $\mathrm{TPR}(l)$.
By varying $l$ from 0 to $N$, two sequences of points $(\mathrm{TPR}(l),\mathrm{FPR}(l)$) and $(\mathrm{Recall}(l),\mathrm{Precision}(l))$ are measured respectively, and the receiver operating characteristic curve and the precision-recall curve are obtained, as shown in Fig.~\ref{fig:validation}(d) and (f). The area under the two curves, denoted as AUROC and AUPR repectively, repensent the reconstruction performance: AUROC(AUPR) ranges from AUC=0.5(AUPR=$P/2N$) for random guessing to AUROC=1(AUPR=1) for perfect reconstructibility.

Because the links of each node
are actually identified separately, the AUROC and AUPR are calculated
for each node, and we use the mean index values over all the nodes to
characterize the reconstruction performance for the whole network.

\subsection{Reconstruction performance affected by network structure and amount of data} \label{sec:numerics}
We test our method by implementing different dynamical processes on
Erd\"os-R\'enyi random (ER)~\cite{erd6s1960}, scale-free
(SF)~\cite{barabasi1999}, small-world (SW)~\cite{watts1998}, and
empirical networks. For network reconstruction, knowledge
about the switching dynamics and network details is not necessary -
only the states of the nodes at different time steps need to be
recorded. See Sec.~1 in Supplementary Materials for computational details.



\begin{figure*}
\centering
\includegraphics[width=0.9\textwidth]{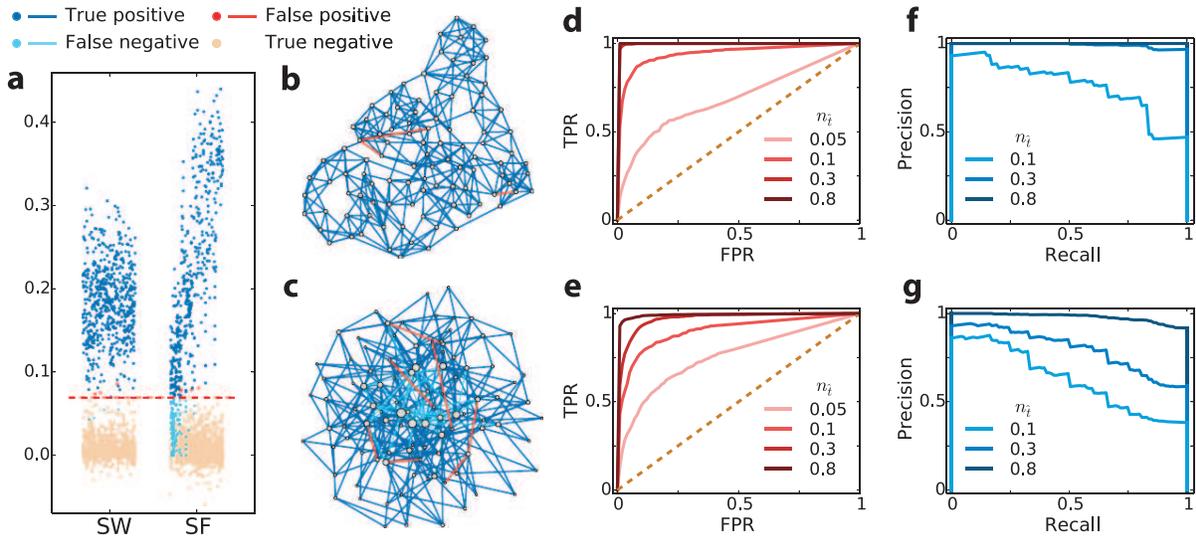}
\caption{{\bf Reconstruction performance.} ({\bf a}) Reconstructed values of the
neighboring vector ${\bf X}_i$ for all nodes in SW and SF networks with
the voter model, where $N = 100$, $\langle k \rangle = 6$, $n_{\hat t}=0.8$ and the length of time series used is $1.5\times 10^4$. The red dashed line represents the threshold for
determining whether a reconstructed value is regarded as representing
an actual link (a value larger than the threshold) or a null link (a value
smaller than the threshold). The correctly reconstructed links (true positive),
falsely reconstructed links (false positive), and missing links (false
negative) are represented by the dark blue, red and light blue points,
respectively, while the yellow points indicate the true negative links.
({\bf b}, {\bf c}) Visualization of the reconstructed SW and SF networks, respectively.
The color legends of the reconstructed links are the same as those in {\bf a}.
There are more missing links (false negative) in the SF network than in the
ER network. ({\bf d}, {\bf e}) ROC curves of reconstructed values for SW and SF networks
for different values of $n_{\hat t}$. ({\bf f}, {\bf g}) PR curves of the reconstructed
values for SW and SF networks for different values of $n_{\hat t}$.}
\label{fig3}
\end{figure*}

\begin{figure}
\centering
\includegraphics[width=0.5\textwidth]{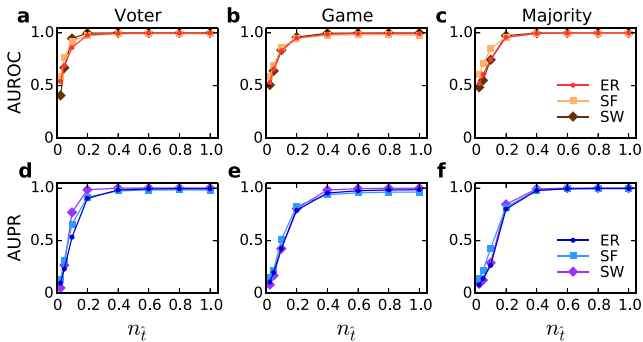}
\caption{{\bf Reconstruction performance with respect to the number of base strings.}
LassoLasso({\bf a,b,c}) AUROC and ({\bf d,e,f}) AUPR as functions of the normalized number of base strings $n_{\hat t}$ for the voter, game and majority model on ER, SF and SW networks. The network size $N=100$ and $\langle k \rangle =6$. The length of time series is $1.5\times 10^4$. Other parameter values of binary-state dynamics are shown in Supplementary Table~S1.}
\label{fig:eq}
\end{figure}

Figure~\ref{fig3} illustrates the reconstruction performance, where
Fig.~\ref{fig3}(a) shows the element values $x_{ij}$ in the reconstructed
neighboring vector ${\bf X}_i$ of all nodes for SW and SF networks
with the voter model. We note that the values of $x_{ij}$ corresponding to
actual links are markedly and distinctly greater than those of null
connections. Setting a cut-off value in the gap between the two groups of
points in Fig.~\ref{fig3}(a), we can separate the actual links from the null
connections, enabling a reconstruction of the whole SW network.
For the SF network, it is difficult to fully reconstruct the neighbors
of the hub nodes, for the following two reasons: (i) in general the
linearization procedure works better for small node degree, as
shown in Fig.~\ref{fig2}; (ii) the lasso based reconstruction requires
smaller data amount and offers better accuracy for sparser vector
${\bf X}_i$ associated with small degree nodes. However, for an SF
network, a vast majority of the nodes in an SF network are not hubs, which
can be precisely reconstructed. The reconstructed SW and SF
networks are shown in Figs.~\ref{fig3}(b) and \ref{fig3}(c), respectively.

\begin{table*}
\linespread{1.7}
\centering
\caption{{\bf AUROC and AUPR measures for various dynamics on a variety of
model and empirical networks.} The parameter values in the dynamical models
are listed in Supplementary Table~S1. The size and mean degree of ER,
SF and SW networks are $N=500$ and $\langle k \rangle =6$, and the length of time series used is $6\times 10^4$.
The length of time series used for empirical networks is $1.5\times 10^4$. }
\scriptsize
\begin{tabular}{c c c c c c c c c}
\hline
AUROC/AUPR & Voter & Kirman & Ising & SIS & Game & Language & Threshold & Majority \\
\hline
ER       & $1.000$/$0.983$ & $0.999$/$0.954$ & $1.000$/$0.982$ & $0.997$/$0.960$ & $0.999$/$0.981$ & $0.995$/$0.934$ & $1.000$/$0.988$ & $1.000$/$0.986$ \\
SF & $0.992$/$0.959$ & $0.985$/$0.920$ & $0.998$/$0.976$ & $0.984$/$0.924$ & $0.988$/$0.951$ & $0.986$/$0.925$ & $0.986$/$0.985$ & $0.999$/$0.980$ \\
SW & $1.000$/$0.988$ & $1.000$/$0.982$ & $1.000$/$0.988$ & $1.000$/$0.988$ & $1.000$/$0.988$ & $1.000$/$0.986$ & $0.994$/$0.979$ & $1.000$/$0.987$ \\
Dolphins & $1.000$/$0.916$ & $0.997$/$0.908$ & $0.999$/$0.911$ & $0.978$/$0.867$ & $0.993$/$0.900$ & $0.985$/$0.870$ & $0.991$/$0.890$ & $1.000$/$0.913$ \\
Football & $0.999$/$0.884$ & $1.000$/$0.898$ & $0.999$/$0.899$ & $0.999$/$0.884$ & $0.996$/$0.882$ & $0.992$/$0.859$ & $0.918$/$0.637$ & $0.999$/$0.896$ \\
Karate & $0.997$/$0.856$ & $0.969$/$0.838$ & $0.981$/$0.836$ & $0.954$/$0.823$ & $0.984$/$0.839$ & $0.960$/$0.803$ & $0.971$/$0.810$ & $0.996$/$0.847$ \\
Leader   & $1.000$/$0.838$ & $0.991$/$0.912$ & $0.991$/$0.823$ & $0.968$/$0.789$ & $0.990$/$0.818$ & $0.966$/$0.780$ & $0.970$/$0.760$ & $0.998$/$0.832$ \\
Polbooks & $0.999$/$0.912$ & $0.991$/$0.829$ & $0.998$/$0.908$ & $0.932$/$0.779$ & $0.986$/$0.888$ & $0.978$/$0.857$ & $0.971$/$0.858$ & $0.999$/$0.913$ \\
Prison & $1.000$/$0.936$ & $0.999$/$0.896$ & $1.000$/$0.935$ & $0.992$/$0.915$ & $0.981$/$0.909$ & $0.991$/$0.909$ & $0.999$/$0.931$ & $1.000$/$0.935$ \\
Santa Fe & $0.998$/$0.967$ & $0.990$/$0.933$ & $1.000$/$0.969$ & $0.982$/$0.937$ & $0.997$/$0.965$ & $0.996$/$0.959$ & $0.994$/$0.961$ & $1.000$/$0.970$ \\
\hline
\end{tabular}
\label{table1}
\end{table*}

To assess how the number of base strings $\hat{t}$ affects the
reconstruction accuracy, we define $n_{\hat t}$ to be the number of
$\hat{t}$ divided by the network size $N$ to quantify the relative amount
of the base strings. As shown in Figs.~\ref{fig3}(d-g), the receiver
operating characteristic (ROC) and the precision-recall (PR) curves
show better performance as $n_{\hat t}$ is increased for both SW and
SF networks, implying that high accuracy can be achieved for reasonably
large values of $n_{\hat t}$.
Fig.~\ref{fig:eq} shows the AUROC and AUPR measures as a function
of $n_{\hat t}$ for different dynamical models
on ER, SW and SF networks. Due to the advantage of the lasso for
sparse vectors, nearly perfect reconstruction is achieved after
$n_{\hat t}$ exceeds a relatively small critical value, e.g., $0.4$.

It is also important to assess how the length of the binary time series
affects the reconstruction accuracy and efficiency. We have calculated
the AUROC and AUPR measures as a function of the normalized time-series
length for various dynamical processes on ER, SF and SW networks (see
(see Supplementary Sec.~2). In general, high reconstruction accuracy can be
achieved for relatively short time series.
We systematically test our method on a variety of model and real networks
in combination with eight binary-state dynamics (Table~\ref{table1}) and
find high values of AUROC and AUPR for all cases.

\begin{figure}
\centering
\includegraphics[width=0.5\textwidth]{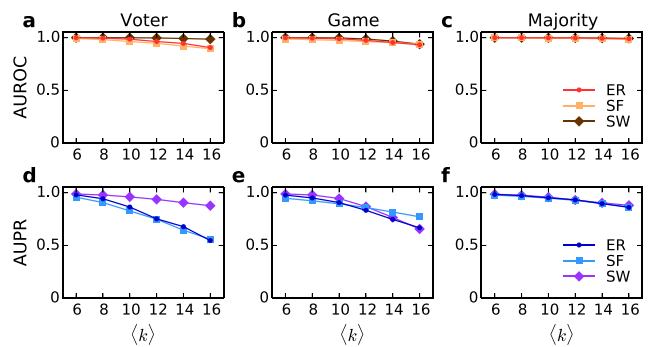}
\caption{{\bf Reconstruction performance affected by average node degree.}
({\bf a,b,c}) AUROC and ({\bf d,e,f}) AUPR as functions of the average node degree $\langle k \rangle$ for the voter, game and majority model on ER, SF and SW networks. The network size $N=500$ and normalized length of time series $n_t=100$. Other parameter values of binary-state dynamics are shown in Supplementary Table~S1.}
\label{fig:k}
\end{figure}
\begin{figure}
\centering
\includegraphics[width=0.5\textwidth]{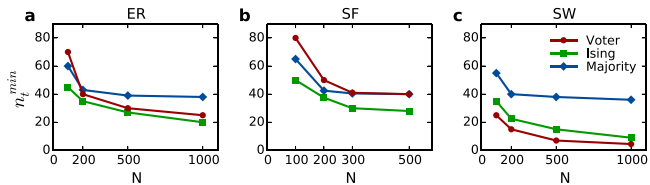}
\caption{{\bf Reconstruction performance affected by network size.}
The minimum normalized length $n_t^{\rm min}$ to acquire at least $0.95$ AUROC and AUPR simultaneously as a function of network size $N$ for the voter, Ising and majority model on ({\bf a}) ER, ({\bf b}) SF and ({\bf c}) SW networks. The mean degree of networks is $6$. Other parameter values of binary-state dynamics are shown in Supplementary Table~S1.}
\label{fig:N}
\end{figure}

We explore the effects of network properties such as the average degree $\langle k\rangle$ and the size $N$ on reconstruction performance.
As shown in Fig.~\ref{fig:k}. The reconstruction accuracy decreases as $\left\langle k \right\rangle$ increases. The main reason for this result is that the low-order approximation in the data-based linearization is better for smaller node degree. Moreover, with the increase of $\left\langle k \right\rangle$, the vector ${\bf X}_i$ to be reconstructed will become denser. Note that it usually requires larger amounts of data to reconstruct a denser signal by using the lasso according to the compressive sensing theory. Thus, in general a network with larger $\left\langle k \right\rangle$ will be more difficult to be reconstructed.
Fig.~\ref{fig:N} shows the minimum normalized length of time series $n_t^{min}$ to acquire at least $0.95$ AUROC and AUPR simultaneously as a function of network size $N$. We see that $n_t^{min}$  decreases as $N$ increases, which is because of network sparsity as well. In general, for the same average node degree $\langle k\rangle$, a network with larger size will be sparser, leading to a sparser vector ${\bf X}_i$. According to the compressive sensing theory, less data are required for reconstructing a sparser ${\bf X}_i$, accounting for the decrease of $n_t^{min}$ with the increase of $N$.
These results indicate that our reconstruction method is scalable and of practical importance for dealing with large real networked systems.

\begin{table*}
\linespread{1.7}
\centering
\caption{{\bf Robustness of reconstruction against noise and missing data.} AUROC and AUPR measures for voter, game, and majority models on ER, SF and SW networks
for measurement noise $n_f= 10\%$ and the fraction of inaccessible nodes
$n_m=30\%$. The network size is $N=500$ and the mean degree is
$\langle k \rangle =6$. The length of the time series used is $6\times 10^4$.
Details of the parameter values in the dynamical models are listed in
Supplementary Table~S1.}
\footnotesize
\begin{tabular}{l | c c c | c c c}
\hline
        &  & $n_f=10\%$ &  &  & $n_m=30\%$ &  \\
\hline
AUROC/AUPR & Voter & Game & Majority & Voter & Game & Majority \\
\hline
ER & $0.995$/$0.938$ & $0.955$/$0.707$ & $0.991$/$0.864$ & $1.000$/$0.985$ & $0.999$/$0.983$ & $1.000$/$0.988$ \\
SF & $0.983$/$0.903$ & $0.954$/$0.800$ & $0.990$/$0.894$ & $0.995$/$0.968$ & $0.991$/$0.957$ & $0.995$/$0.984$ \\
SW & $1.000$/$0.984$ & $0.976$/$0.741$ & $0.994$/$0.874$ & $1.000$/$0.988$ & $1.000$/$0.988$ & $1.000$/$0.988$ \\
\hline
\end{tabular}
\label{table2}
\end{table*}

\subsection{Robustness of reconstruction against noise and missing data}

In real applications, time series are often contaminated by noise
and the data from certain nodes may be lost or inaccessible. To address
these practical issues, we test the robustness of our method. Specifically,
we instill noise into the time series by randomly flipping a fraction $n_f$
of binary states and assume a fraction $n_m$ of nodes are inaccessible.
The results are shown in Table~\ref{table2}, where voter, game, and majority
models are used as examples of linear, nonlinear and piecewise dynamics,
respectively. Strikingly, we obtain high values of AUROC and AUPR even
in presence of $10\%$ measurement noise or $30\%$ inaccessible nodes,
providing strong evidence for the robustness of our framework against
noise and missing data. More detailed characterization associated with
the results in Table~\ref{table2}, i.e., AUROC and AUPR as functions of
$n_f$ and $n_m$, are provided in Supplementary Sec.~3.

\section{Discussion} \label{sec:discussion}


Reconstructing the topological structure and dynamics of complex systems from data
is a central issue in both network science and engineering community~\cite{CCPGP:2013,GdiBLC:2003,Timme:2007,
BL:2007,CMN:2008,RWLL:2010,LP:2011,HKKN:2011}.
A framework~\cite{WYLKG:2011,WYLKH:2011,WLGY:2011} of network
reconstruction is based on compressive sensing~\cite{CRT:2006a,
CRT:2006b,Donoho:2006,Baraniuk:2007,CW:2008,Romberg:2008},
a sparse signal recovery method developed in applied mathematics
and engineering signal processing. A recent work~\cite{SWFDL:2014}
also demonstrated that compressive sensing can be exploited for
network reconstruction in situations where the available time series
are polarized (binary), e.g., virus spreading and information
diffusion in social and computer networks. While the structure of
the virus propagation network and the spreading sources can be obtained,
the method is unable to predict the network dynamical systems that
generate the binary data.

The contribution of this paper is a general framework to solve the
challenging problem of reconstructing complex networks hosting
binary-state dynamics, based only on time series without any
knowledge of the network structure and the switching functions that
generate the binary data. The key to our success is the formulation
of a universal data-based linearization method, which is powerful
for reconstructing the neighborhood of nodes for any type of nodal
dynamics: linear, nonlinear, discontinuous, or stochastic. The natural
sparsity of real complex networks allows us to address the local
reconstruction as a sparse signal reconstruction problem that can be
solved by employing the lasso, a convex optimization method, from small
amounts of binary data. The optimization is robust against measurement
noise and missing data. Once the neighborhoods of all nodes have been
reconstructed, the whole network can be mapped out by assembling all
the local structures and making adjustments to ensure consistency.
We have validated our framework using a variety of binary-state dynamical
models on a number of model and real complex networks. High reconstruction
accuracy has been obtained for all cases, even for relatively small amounts
of binary data contaminated by noise and when partial data are lost. These
results suggest the practical applicability of our framework.

While our framework potentially offers a general, completely data driven
approach to reconstructing binary dynamical processes on complex networks,
there are still challenges. For example, our framework can deal with
various types of switching functions underlying the binary-state dynamics,
but in its present form the framework is not applicable to non-monotonous
functions or non-Markovian type of dynamics. Especially, when the switching
functions are not monotonous, the data-based linearization would fail due to
the violation of the one-to-one correspondence between the switching
probability and the number of active neighbors. For non-Markovian dynamics,
the merging procedure inherent in our method would fail. To predict the
interaction strength among nodes presents another challenge, especially
where noise is present and there is missing data. The results reported
in this paper suggest strongly that our present framework can serve as
a starting point to meet the challenges, eventually leading to
a complete and universally applicable solution to the inverse problem
of binary network structure and dynamics.


\begin{thebibliography}{10}

\bibitem{barrat2008}
\bibinfo{author}{Barrat, A.}, \bibinfo{author}{Barthelemy, M.} \&
  \bibinfo{author}{Vespignani, A.}
\newblock \emph{\bibinfo{title}{Dynamical Processes on Complex Networks}}
  (\bibinfo{publisher}{Cambridge University Press}, \bibinfo{year}{2008}).

\bibitem{kumar2010}
\bibinfo{author}{Kumar, A.}, \bibinfo{author}{Rotter, S.} \&
  \bibinfo{author}{Aertsen, A.}
\newblock \bibinfo{title}{Spiking activity propagation in neuronal networks:
  reconciling different perspectives on neural coding}.
\newblock \emph{\bibinfo{journal}{Nature Rev. Neuro.}}
  \textbf{\bibinfo{volume}{11}}, \bibinfo{pages}{615--627}
  (\bibinfo{year}{2010}).

\bibitem{game}
\bibinfo{author}{Szab{\'o}, G.} \& \bibinfo{author}{Fath, G.}
\newblock \bibinfo{title}{Evolutionary games on graphs}.
\newblock \emph{\bibinfo{journal}{Phys. Rep.}} \textbf{\bibinfo{volume}{446}},
  \bibinfo{pages}{97--216} (\bibinfo{year}{2007}).

\bibitem{pastor2015}
\bibinfo{author}{Pastor-Satorras, R.}, \bibinfo{author}{Castellano, C.},
  \bibinfo{author}{Van~Mieghem, P.} \& \bibinfo{author}{Vespignani, A.}
\newblock \bibinfo{title}{Epidemic processes in complex networks}.
\newblock \emph{\bibinfo{journal}{Rev. Mod. Phys.}}
  \textbf{\bibinfo{volume}{87}}, \bibinfo{pages}{925--979}
  (\bibinfo{year}{2015}).

\bibitem{shao2009}
\bibinfo{author}{Shao, J.}, \bibinfo{author}{Havlin, S.} \&
  \bibinfo{author}{Stanley, H.~E.}
\newblock \bibinfo{title}{Dynamic opinion model and invasion percolation}.
\newblock \emph{\bibinfo{journal}{Phys. Rev. Lett.}}
  \textbf{\bibinfo{volume}{103}}, \bibinfo{pages}{018701}
  (\bibinfo{year}{2009}).

\bibitem{granell2013}
\bibinfo{author}{Granell, C.}, \bibinfo{author}{G{\'o}mez, S.} \&
  \bibinfo{author}{Arenas, A.}
\newblock \bibinfo{title}{Dynamical interplay between awareness and epidemic
  spreading in multiplex networks}.
\newblock \emph{\bibinfo{journal}{Phys. Rev. Lett.}}
  \textbf{\bibinfo{volume}{111}}, \bibinfo{pages}{128701}
  (\bibinfo{year}{2013}).

\bibitem{santos2005}
\bibinfo{author}{Santos, F.~C.} \& \bibinfo{author}{Pacheco, J.~M.}
\newblock \bibinfo{title}{Scale-free networks provide a unifying framework for
  the emergence of cooperation}.
\newblock \emph{\bibinfo{journal}{Phys. Rev. Lett.}}
  \textbf{\bibinfo{volume}{95}}, \bibinfo{pages}{098104}
  (\bibinfo{year}{2005}).

\bibitem{koseska2013}
\bibinfo{author}{Koseska, A.}, \bibinfo{author}{Volkov, E.} \&
  \bibinfo{author}{Kurths, J.}
\newblock \bibinfo{title}{Oscillation quenching mechanisms: Amplitude vs.
  oscillation death}.
\newblock \emph{\bibinfo{journal}{Phys. Rep.}} \textbf{\bibinfo{volume}{531}},
  \bibinfo{pages}{173--199} (\bibinfo{year}{2013}).

\bibitem{buldyrev2010}
\bibinfo{author}{Buldyrev, S.~V.}, \bibinfo{author}{Parshani, R.},
  \bibinfo{author}{Paul, G.}, \bibinfo{author}{Stanley, H.~E.} \&
  \bibinfo{author}{Havlin, S.}
\newblock \bibinfo{title}{Catastrophic cascade of failures in interdependent
  networks}.
\newblock \emph{\bibinfo{journal}{Nature}} \textbf{\bibinfo{volume}{464}},
  \bibinfo{pages}{1025--1028} (\bibinfo{year}{2010}).

\bibitem{galbiati2013}
\bibinfo{author}{Galbiati, M.}, \bibinfo{author}{Delpini, D.} \&
  \bibinfo{author}{Battiston, S.}
\newblock \bibinfo{title}{The power to control}.
\newblock \emph{\bibinfo{journal}{Nature Phys.}} \textbf{\bibinfo{volume}{9}},
  \bibinfo{pages}{126--128} (\bibinfo{year}{2013}).

\bibitem{balcan2011}
\bibinfo{author}{Balcan, D.} \& \bibinfo{author}{Vespignani, A.}
\newblock \bibinfo{title}{Phase transitions in contagion processes mediated by
  recurrent mobility patterns}.
\newblock \emph{\bibinfo{journal}{Nature Phys.}} \textbf{\bibinfo{volume}{7}},
  \bibinfo{pages}{581--586} (\bibinfo{year}{2011}).

\bibitem{newman2010}
\bibinfo{author}{Newman, M.}
\newblock \emph{\bibinfo{title}{Networks: An Introduction}}
  (\bibinfo{publisher}{Oxford University Press}, \bibinfo{year}{2010}).

\bibitem{voter}
\bibinfo{author}{Sood, V.} \& \bibinfo{author}{Redner, S.}
\newblock \bibinfo{title}{Voter model on heterogeneous graphs}.
\newblock \emph{\bibinfo{journal}{Phys. Rev. Lett.}}
  \textbf{\bibinfo{volume}{94}}, \bibinfo{pages}{178701}
  (\bibinfo{year}{2005}).

\bibitem{sis}
\bibinfo{author}{Pastor-Satorras, R.} \& \bibinfo{author}{Vespignani, A.}
\newblock \bibinfo{title}{Epidemic spreading in scale-free networks}.
\newblock \emph{\bibinfo{journal}{Phys. Rev. Lett.}}
  \textbf{\bibinfo{volume}{86}}, \bibinfo{pages}{3200} (\bibinfo{year}{2001}).

\bibitem{castellano2009}
\bibinfo{author}{Castellano, C.}, \bibinfo{author}{Fortunato, S.} \&
  \bibinfo{author}{Loreto, V.}
\newblock \bibinfo{title}{Statistical physics of social dynamics}.
\newblock \emph{\bibinfo{journal}{Rev. Mod. Phys.}}
  \textbf{\bibinfo{volume}{81}}, \bibinfo{pages}{591} (\bibinfo{year}{2009}).

\bibitem{bashan2013}
\bibinfo{author}{Bashan, A.}, \bibinfo{author}{Berezin, Y.},
  \bibinfo{author}{Buldyrev, S.~V.} \& \bibinfo{author}{Havlin, S.}
\newblock \bibinfo{title}{The extreme vulnerability of interdependent spatially
  embedded networks}.
\newblock \emph{\bibinfo{journal}{Nature Phys.}} \textbf{\bibinfo{volume}{9}},
  \bibinfo{pages}{667--672} (\bibinfo{year}{2013}).

\bibitem{ising}
\bibinfo{author}{Krapivsky, P.~L.}, \bibinfo{author}{Redner, S.} \&
  \bibinfo{author}{Ben-Naim, E.}
\newblock \emph{\bibinfo{title}{A Kinetic View of Statistical Physics}}
  (\bibinfo{publisher}{Cambridge University Press}, \bibinfo{year}{2010}).

\bibitem{santos2008}
\bibinfo{author}{Santos, F.~C.}, \bibinfo{author}{Santos, M.~D.} \&
  \bibinfo{author}{Pacheco, J.~M.}
\newblock \bibinfo{title}{Social diversity promotes the emergence of
  cooperation in public goods games}.
\newblock \emph{\bibinfo{journal}{Nature}} \textbf{\bibinfo{volume}{454}},
  \bibinfo{pages}{213--216} (\bibinfo{year}{2008}).

\bibitem{gleeson2013}
\bibinfo{author}{Gleeson, J.~P.}
\newblock \bibinfo{title}{Binary-state dynamics on complex networks: pair
  approximation and beyond}.
\newblock \emph{\bibinfo{journal}{Phys. Rev. X}} \textbf{\bibinfo{volume}{3}},
  \bibinfo{pages}{021004} (\bibinfo{year}{2013}).

\bibitem{boccaletti2006}
\bibinfo{author}{Boccaletti, S.}, \bibinfo{author}{Latora, V.},
  \bibinfo{author}{Moreno, Y.}, \bibinfo{author}{Chavez, M.} \&
  \bibinfo{author}{Hwang, D.-U.}
\newblock \bibinfo{title}{Complex networks: Structure and dynamics}.
\newblock \emph{\bibinfo{journal}{Phys. Rep.}} \textbf{\bibinfo{volume}{424}},
  \bibinfo{pages}{175--308} (\bibinfo{year}{2006}).

\bibitem{barabasi2011}
\bibinfo{author}{Barab{\'a}si, A.-L.}
\newblock \bibinfo{title}{The network takeover}.
\newblock \emph{\bibinfo{journal}{Nature Phys.}} \textbf{\bibinfo{volume}{8}},
  \bibinfo{pages}{14} (\bibinfo{year}{2011}).

\bibitem{GdiBLC:2003}
\bibinfo{author}{Gardner, T.~S.}, \bibinfo{author}{di~Bernardo, D.},
  \bibinfo{author}{Lorenz, D.} \& \bibinfo{author}{Collins, J.~J.}
\newblock \bibinfo{title}{Inferring genetic networks and identifying compound
  mode of action via expression profiling}.
\newblock \emph{\bibinfo{journal}{Science}} \textbf{\bibinfo{volume}{301}},
  \bibinfo{pages}{102--105} (\bibinfo{year}{2003}).

\bibitem{friedman2004}
\bibinfo{author}{Friedman, N.}
\newblock \bibinfo{title}{Inferring cellular networks using probabilistic
  graphical models}.
\newblock \emph{\bibinfo{journal}{Science}} \textbf{\bibinfo{volume}{303}},
  \bibinfo{pages}{799--805} (\bibinfo{year}{2004}).

\bibitem{Timme:2007}
\bibinfo{author}{Timme, M.}
\newblock \bibinfo{title}{Revealing network connectivity from response
  dynamics}.
\newblock \emph{\bibinfo{journal}{Phys. Rev. Lett.}}
  \textbf{\bibinfo{volume}{98}}, \bibinfo{pages}{224101}
  (\bibinfo{year}{2007}).

\bibitem{CMN:2008}
\bibinfo{author}{Clauset, A.}, \bibinfo{author}{Moore, C.} \&
  \bibinfo{author}{Newman, M. E.~J.}
\newblock \bibinfo{title}{Hierarchical structure and the prediction of missing
  links in networks}.
\newblock \emph{\bibinfo{journal}{Nature}} \textbf{\bibinfo{volume}{453}},
  \bibinfo{pages}{98--101} (\bibinfo{year}{2008}).

\bibitem{guo2008}
\bibinfo{author}{Guo, S.}, \bibinfo{author}{Wu, J.}, \bibinfo{author}{Ding, M.}
  \& \bibinfo{author}{Feng, J.}
\newblock \bibinfo{title}{Uncovering interactions in the frequency domain}.
\newblock \emph{\bibinfo{journal}{PLoS Comput. Biol.}}
  \textbf{\bibinfo{volume}{4}}, \bibinfo{pages}{e1000087}
  (\bibinfo{year}{2008}).

\bibitem{RWLL:2010}
\bibinfo{author}{Ren, J.}, \bibinfo{author}{Wang, W.-X.}, \bibinfo{author}{Li,
  B.} \& \bibinfo{author}{Lai, Y.-C.}
\newblock \bibinfo{title}{Noise bridges dynamical correlation and topology in
  coupled oscillator networks}.
\newblock \emph{\bibinfo{journal}{Phys. Rev. Lett.}}
  \textbf{\bibinfo{volume}{104}}, \bibinfo{pages}{058701}
  (\bibinfo{year}{2010}).

\bibitem{HKKN:2011}
\bibinfo{author}{Hempel, S.}, \bibinfo{author}{Koseska, A.},
  \bibinfo{author}{Kurths, J.} \& \bibinfo{author}{Nikoloski, Z.}
\newblock \bibinfo{title}{Inner composition alignment for inferring directed
  networks from short time series}.
\newblock \emph{\bibinfo{journal}{Phys. Rev. Lett.}}
  \textbf{\bibinfo{volume}{107}}, \bibinfo{pages}{054101}
  (\bibinfo{year}{2011}).

\bibitem{WLGY:2011}
\bibinfo{author}{Wang, W.-X.}, \bibinfo{author}{Lai, Y.-C.},
  \bibinfo{author}{Grebogi, C.} \& \bibinfo{author}{Ye, J.-P.}
\newblock \bibinfo{title}{Network reconstruction based on evolutionary-game
  data via compressive sensing}.
\newblock \emph{\bibinfo{journal}{Phys. Rev. X}} \textbf{\bibinfo{volume}{1}},
  \bibinfo{pages}{021021} (\bibinfo{year}{2011}).

\bibitem{barzel2013}
\bibinfo{author}{Barzel, B.} \& \bibinfo{author}{Barab{\'a}si, A.-L.}
\newblock \bibinfo{title}{Network link prediction by global silencing of
  indirect correlations}.
\newblock \emph{\bibinfo{journal}{Nature Biotechnol.}}
  \textbf{\bibinfo{volume}{31}}, \bibinfo{pages}{720--725}
  (\bibinfo{year}{2013}).

\bibitem{feizi2013}
\bibinfo{author}{Feizi, S.}, \bibinfo{author}{Marbach, D.},
  \bibinfo{author}{M{\'e}dard, M.} \& \bibinfo{author}{Kellis, M.}
\newblock \bibinfo{title}{Network deconvolution as a general method to
  distinguish direct dependencies in networks}.
\newblock \emph{\bibinfo{journal}{Nature Biotechnol.}}
  \textbf{\bibinfo{volume}{31}}, \bibinfo{pages}{726--733}
  (\bibinfo{year}{2013}).

\bibitem{CCPGP:2013}
\bibinfo{author}{Caldarelli, G.}, \bibinfo{author}{Chessa, A.},
  \bibinfo{author}{Pammolli, F.}, \bibinfo{author}{Gabrielli, A.} \&
  \bibinfo{author}{Puliga, M.}
\newblock \bibinfo{title}{Reconstructing a credit network}.
\newblock \emph{\bibinfo{journal}{Nature Phys.}} \textbf{\bibinfo{volume}{9}},
  \bibinfo{pages}{125--126} (\bibinfo{year}{2013}).

\bibitem{SWFDL:2014}
\bibinfo{author}{Shen, Z.-S.}, \bibinfo{author}{Wang, W.-X.},
  \bibinfo{author}{Fan, Y.}, \bibinfo{author}{Di, Z.-R.} \&
  \bibinfo{author}{Lai, Y.-C.}
\newblock \bibinfo{title}{Reconstructing propagation networks with natural
  diversity and identifying hidden source}.
\newblock \emph{\bibinfo{journal}{Nature Commun.}}
  \textbf{\bibinfo{volume}{5}}, \bibinfo{pages}{4323} (\bibinfo{year}{2014}).

\bibitem{han2015}
\bibinfo{author}{Han, X.}, \bibinfo{author}{Shen, Z.}, \bibinfo{author}{Wang,
  W.-X.} \& \bibinfo{author}{Di, Z.}
\newblock \bibinfo{title}{Robust reconstruction of complex networks from sparse
  data}.
\newblock \emph{\bibinfo{journal}{Phys. Rev. Lett.}}
  \textbf{\bibinfo{volume}{114}}, \bibinfo{pages}{028701}
  (\bibinfo{year}{2015}).

\bibitem{CRT:2006a}
\bibinfo{author}{Cand{\`e}s, E.~J.}, \bibinfo{author}{Romberg, J.~K.} \&
  \bibinfo{author}{Tao, T.}
\newblock \bibinfo{title}{Robust uncertainty principles: Exact signal
  reconstruction from highly incomplete frequency information}.
\newblock \emph{\bibinfo{journal}{IEEE Trans. Info. Theo.}}
  \textbf{\bibinfo{volume}{52}}, \bibinfo{pages}{489--509}
  (\bibinfo{year}{2006}).

\bibitem{CRT:2006b}
\bibinfo{author}{Candes, E.~J.}, \bibinfo{author}{Romberg, J.~K.} \&
  \bibinfo{author}{Tao, T.}
\newblock \bibinfo{title}{Stable signal recovery from incomplete and inaccurate
  measurements}.
\newblock \emph{\bibinfo{journal}{Commun. Pure Appl. Math.}}
  \textbf{\bibinfo{volume}{59}}, \bibinfo{pages}{1207--1223}
  (\bibinfo{year}{2006}).

\bibitem{Donoho:2006}
\bibinfo{author}{Donoho, D.~L.}
\newblock \bibinfo{title}{Compressed sensing}.
\newblock \emph{\bibinfo{journal}{IEEE Trans. Info. Theo.}}
  \textbf{\bibinfo{volume}{52}}, \bibinfo{pages}{1289--1306}
  (\bibinfo{year}{2006}).

\bibitem{Baraniuk:2007}
\bibinfo{author}{Baraniuk, R.~G.}
\newblock \bibinfo{title}{Compressive sensing}.
\newblock \emph{\bibinfo{journal}{IEEE Sig. Proc. Mag.}}
  \textbf{\bibinfo{volume}{24}}, \bibinfo{pages}{118--121}
  (\bibinfo{year}{2007}).

\bibitem{CW:2008}
\bibinfo{author}{Cand{\`e}s, E.~J.} \& \bibinfo{author}{Wakin, M.~B.}
\newblock \bibinfo{title}{An introduction to compressive sampling}.
\newblock \emph{\bibinfo{journal}{IEEE Sig. Proc. Mag.}}
  \textbf{\bibinfo{volume}{25}}, \bibinfo{pages}{21--30}
  (\bibinfo{year}{2008}).

\bibitem{Romberg:2008}
\bibinfo{author}{Romberg, J.}
\newblock \bibinfo{title}{Imaging via compressive sampling}.
\newblock \emph{\bibinfo{journal}{IEEE Sig. Proc. Mag.}}
  \textbf{\bibinfo{volume}{25}}, \bibinfo{pages}{14--20}
  (\bibinfo{year}{2008}).

\bibitem{lasso}
\bibinfo{author}{Hastie, T.}, \bibinfo{author}{Tibshirani, R.},
  \bibinfo{author}{Friedman, J.} \& \bibinfo{author}{Franklin, J.}
\newblock \bibinfo{title}{The elements of statistical learning: data mining,
  inference and prediction}.
\newblock \emph{\bibinfo{journal}{Math. Intell.}}
  \textbf{\bibinfo{volume}{27}}, \bibinfo{pages}{83--85}
  (\bibinfo{year}{2005}).

\bibitem{liu2011}
\bibinfo{author}{Liu, Y.-Y.}, \bibinfo{author}{Slotine, J.-J.} \&
  \bibinfo{author}{Barab{\'a}si, A.-L.}
\newblock \bibinfo{title}{Controllability of complex networks}.
\newblock \emph{\bibinfo{journal}{Nature}} \textbf{\bibinfo{volume}{473}},
  \bibinfo{pages}{167--173} (\bibinfo{year}{2011}).

\bibitem{nepusz2012}
\bibinfo{author}{Nepusz, T.} \& \bibinfo{author}{Vicsek, T.}
\newblock \bibinfo{title}{Controlling edge dynamics in complex networks}.
\newblock \emph{\bibinfo{journal}{Nature Phys.}} \textbf{\bibinfo{volume}{8}},
  \bibinfo{pages}{568--573} (\bibinfo{year}{2012}).

\bibitem{yan2012}
\bibinfo{author}{Yan, G.}, \bibinfo{author}{Ren, J.}, \bibinfo{author}{Lai,
  Y.-C.}, \bibinfo{author}{Lai, C.-H.} \& \bibinfo{author}{Li, B.}
\newblock \bibinfo{title}{Controlling complex networks: How much energy is
  needed?}
\newblock \emph{\bibinfo{journal}{Phys. Rev. Lett.}}
  \textbf{\bibinfo{volume}{108}}, \bibinfo{pages}{218703}
  (\bibinfo{year}{2012}).

\bibitem{yuan2013}
\bibinfo{author}{Yuan, Z.}, \bibinfo{author}{Zhao, C.}, \bibinfo{author}{Di,
  Z.}, \bibinfo{author}{Wang, W.-X.} \& \bibinfo{author}{Lai, Y.-C.}
\newblock \bibinfo{title}{Exact controllability of complex networks}.
\newblock \emph{\bibinfo{journal}{Nature Commun.}} \textbf{\bibinfo{volume}{4}}
  (\bibinfo{year}{2013}).

\bibitem{RR:2014}
\bibinfo{author}{Ruths, J.} \& \bibinfo{author}{Ruths, D.}
\newblock \bibinfo{title}{Control profiles of complex networks}.
\newblock \emph{\bibinfo{journal}{Science}} \textbf{\bibinfo{volume}{343}},
  \bibinfo{pages}{1373--1376} (\bibinfo{year}{2014}).

\bibitem{Wuchty:2014}
\bibinfo{author}{Wuchty, S.}
\newblock \bibinfo{title}{Controllability in protein interaction networks}.
\newblock \emph{\bibinfo{journal}{Proc. Natl. Acad. Sci. (USA)}}
  \textbf{\bibinfo{volume}{111}}, \bibinfo{pages}{7156--7160}
  (\bibinfo{year}{2014}).

\bibitem{erd6s1960}
\bibinfo{author}{Erd{\"o}s, P.} \& \bibinfo{author}{R{\'e}nyi, A.}
\newblock \bibinfo{title}{On the evolution of random graphs}.
\newblock \emph{\bibinfo{journal}{Publ. Math. Inst. Hungar. Acad. Sci}}
  \textbf{\bibinfo{volume}{5}}, \bibinfo{pages}{17--61} (\bibinfo{year}{1960}).

\bibitem{barabasi1999}
\bibinfo{author}{Barab{\'a}si, A.-L.} \& \bibinfo{author}{Albert, R.}
\newblock \bibinfo{title}{Emergence of scaling in random networks}.
\newblock \emph{\bibinfo{journal}{Science}} \textbf{\bibinfo{volume}{286}},
  \bibinfo{pages}{509--512} (\bibinfo{year}{1999}).

\bibitem{watts1998}
\bibinfo{author}{Watts, D.~J.} \& \bibinfo{author}{Strogatz, S.~H.}
\newblock \bibinfo{title}{Collective dynamics of ¡®small-world¡¯networks}.
\newblock \emph{\bibinfo{journal}{Nature}} \textbf{\bibinfo{volume}{393}},
  \bibinfo{pages}{440--442} (\bibinfo{year}{1998}).

\bibitem{davis2006}
\bibinfo{author}{Davis, J.} \& \bibinfo{author}{Goadrich, M.}
\newblock \bibinfo{title}{The relationship between precision-recall and roc
  curves}.
\newblock In \emph{\bibinfo{booktitle}{Proceedings of the 23rd international
  conference on Machine learning}}, \bibinfo{pages}{233--240}
  (\bibinfo{organization}{ACM}, \bibinfo{year}{2006}).

\bibitem{BL:2007}
\bibinfo{author}{Bongard, J.} \& \bibinfo{author}{Lipson, H.}
\newblock \bibinfo{title}{Automated reverse engineering of nonlinear dynamical
  systems}.
\newblock \emph{\bibinfo{journal}{Proc. Natl. Acad. Sci. (USA)}}
  \textbf{\bibinfo{volume}{104}}, \bibinfo{pages}{9943--9948}
  (\bibinfo{year}{2007}).

\bibitem{LP:2011}
\bibinfo{author}{Levnaji\'{c}, Z.} \& \bibinfo{author}{Pikovsky, A.}
\newblock \bibinfo{title}{Network reconstruction from random phase resetting}.
\newblock \emph{\bibinfo{journal}{Phys. Rev. Lett.}}
  \textbf{\bibinfo{volume}{107}}, \bibinfo{pages}{034101}
  (\bibinfo{year}{2011}).

\bibitem{WYLKG:2011}
\bibinfo{author}{Wang, W.-X.}, \bibinfo{author}{Yang, R.},
  \bibinfo{author}{Lai, Y.-C.}, \bibinfo{author}{Kovanis, V.} \&
  \bibinfo{author}{Grebogi, C.}
\newblock \bibinfo{title}{Predicting catastrophes in nonlinear dynamical
  systems by compressive sensing}.
\newblock \emph{\bibinfo{journal}{Phys. Rev. Lett.}}
  \textbf{\bibinfo{volume}{106}}, \bibinfo{pages}{154101}
  (\bibinfo{year}{2011}).

\bibitem{WYLKH:2011}
\bibinfo{author}{Wang, W.-X.}, \bibinfo{author}{Yang, R.},
  \bibinfo{author}{Lai, Y.-C.}, \bibinfo{author}{Kovanis, V.} \&
  \bibinfo{author}{Harrison, M. A.~F.}
\newblock \bibinfo{title}{Time-series--based prediction of complex oscillator
  networks via compressive sensing}.
\newblock \emph{\bibinfo{journal}{EPL}} \textbf{\bibinfo{volume}{94}},
  \bibinfo{pages}{48006} (\bibinfo{year}{2011}).

\bibitem{majority}
\bibinfo{author}{de~Oliveira, M.~J.}
\newblock \bibinfo{title}{Isotropic majority-vote model on a square lattice}.
\newblock \emph{\bibinfo{journal}{J. Stat. Phys.}}
  \textbf{\bibinfo{volume}{66}}, \bibinfo{pages}{273--281}
  (\bibinfo{year}{1992}).

\bibitem{kirman}
\bibinfo{author}{Kirman, A.}
\newblock \bibinfo{title}{Ants, rationality, and recruitment}.
\newblock \emph{\bibinfo{journal}{Quart. J. Econo.}} \bibinfo{pages}{137--156}
  (\bibinfo{year}{1993}).

\bibitem{language}
\bibinfo{author}{Abrams, D.~M.} \& \bibinfo{author}{Strogatz, S.~H.}
\newblock \bibinfo{title}{Linguistics: Modelling the dynamics of language
  death}.
\newblock \emph{\bibinfo{journal}{Nature}} \textbf{\bibinfo{volume}{424}},
  \bibinfo{pages}{900--900} (\bibinfo{year}{2003}).

\bibitem{threshold}
\bibinfo{author}{Granovetter, M.}
\newblock \bibinfo{title}{Threshold models of collective behavior}.
\newblock \emph{\bibinfo{journal}{Ame. J. Socio.}} \bibinfo{pages}{1420--1443}
  (\bibinfo{year}{1978}).

\bibitem{glauber}
\bibinfo{author}{Glauber, R.~J.}
\newblock \bibinfo{title}{Time-dependent statistics of the ising model}.
\newblock \emph{\bibinfo{journal}{J. Math. Phys.}}
  \textbf{\bibinfo{volume}{4}}, \bibinfo{pages}{294--307}
  (\bibinfo{year}{1963}).

\bibitem{lasso_python}
\bibinfo{author}{Pedregosa, F.} \emph{et~al.}
\newblock \bibinfo{title}{Scikit-learn: Machine learning in python}.
\newblock \emph{\bibinfo{journal}{J. Mach. Learn. Res.}}
  \textbf{\bibinfo{volume}{12}}, \bibinfo{pages}{2825--2830}
  (\bibinfo{year}{2011}).

\end{thebibliography}

\appendix

\section{Description of used Binary-state Dynamics}
The voter model~\cite{voter} assumes that a node randomly chooses and
then adopts one of its neighbors' state at each time step. The total
number of neighbors is its degree $k$, of which $m$ are active,
i.e., they are in state $1$. The probabilities that the node will become
active and inactive are $m/k$ and $(k-m)/k$, respectively. In the
majority-voter model~\cite{majority}, a node tends to align with the
majority state of its neighbors, and the probability of misalignment
is $Q$.

In the Kirman's ant colony model~\cite{kirman}, a node switches from
state $0$ to $1$ with the probability $F_{k,m}=c_1+dm$ (with $m$ being
the number of active neighbors) and the rate of transition from $1$ to
$0$ is $R_{k,m}=c_2+d(k-m)$, where the parameters $c_1$ and $c_2$
quantify the individual action that is independent of the states
of the neighbors and $d$ characterizes the action of copying from
neighbors' state.

The Ising model~\cite{ising} is a classic paradigm to study
ferromagnetism at the microscopic level of spins. In the model, a
node can assume either one of the two states: spin-up or spin-down.
Switching in the state occurs with the probability determined by
minimizing the energy (Hamiltonian) of the system. In our study, we
chose the transition rates according to the Glauber
dynamics~\cite{glauber}, as shown in Table~\ref{tab:dynamics}, where
the parameter $\beta$ quantifies the combining effect of temperature
and the ferromagnetic-interaction parameter.

The SIS model~\cite{sis} describes the epidemic process of disease
spreading with infection and recovery. Each susceptible individual
contracts the disease from each of its infected neighbors at the
rate $\lambda$, so at each time step a susceptible node with $m$
infected neighbors has the probability $(1-\lambda )^m$ of remaining
susceptible. The infection rate is then $1-(1- \lambda )^m$. The
recovery rate of an infected node is $\mu$ at each time step.

The game model~\cite{game} originates from the evolutionary game
theory. In a network, each node is a player, and the
two states means that the player can take on two different strategies.
A player plays with each of his/her neighbors using one chosen
strategy at each time step. The profit of a rational player $i$,
when playing with a neighbor $j$, is characterized by the payoff matrix
$\begin{array}{@{}r@{}c@{}c@{}c@{}c@{}l@{}}
    & s_1 & s_2 \\
    \left.\begin{array}
    {c} s_1 \\s_2 \end{array}\right(
                    & \begin{array}{c} a \\ c\end{array}
                    & \begin{array}{c} b \\ d \end{array}
                          & \left)\begin{array}{c} \\ \\  \end{array}\right.
\end{array}$
where $a, b, c$ and $d$ are parameters. Different games can be generated
by adjusting $a, b, c$ and $b$. The payoff of a player is the sum of
profit from playing game with all its neighbors. A player switches
the strategy with a probability that depends on the payoff it may
gain in the next round under the current circumstance by switching its strategy, as illustrated
in Table~\ref{tab:dynamics}, where the parameter $\alpha$ qualifies the
willingness for an individual to change its strategy according to
those of its neighbors, and $\beta$ is associated with the effect
of the expected payoff.

For the language model~\cite{language}, the two states denote
two different language choices of a person. Transition from the
primary language to the secondary occurs with the probability
that is proportional to the fraction of speakers in the neighbors
with the power $\alpha$, multiplied by the parameter $s$ (or $1-s$)
according to the respective language.

The threshold model~\cite{threshold} is a deterministic model, where
for each node a certain threshold $M_k$ is set which can be, for
example, a function of the node's degree. At each time step, a node
becomes active if the number $m$ of its active neighbors exceeds the
threshold $M_k$, and no recovery transformation is permitted.

\section*{Acknowledgements}

W.-X.W. was supported by NSFC under Grant No. 11105011, CNNSF under
Grant No. 61074116 and the Fundamental Research Funds for the
Central Universities. Y.-C.L. was supported by ARO under Grant
W911NF-14-1-0504.

\section*{Author contributions}

W.-X.W. designed research; J.L. and Z.S. performed research;
all analyzed data; J.L., W.-X.W. and Y.-C.L. wrote the paper; all edited the paper.

\section*{Additional information}

Competing financial interests: The authors declare no competing
financial interests.

\newpage

\end{document}